\documentclass{article}
\usepackage{fleqn}
\usepackage{psfig}
\usepackage{epsfig}
\usepackage{graphicx}
\usepackage{amsmath}

\large

\def\beq{\begin{equation}}
\def\eeq{\end{equation}}
\def\bea{\begin{eqnarray}}
\def\eea{\end{eqnarray}}
\def\bq{\begin{quote}}
\def\eq{\end{quote}}

\def\gappeq{\mathrel{\rlap {\raise.5ex\hbox{$>$}}
{\lower.5ex\hbox{$\sim$}}}}
\def\lappeq{\mathrel{\rlap{\raise.5ex\hbox{$<$}}
{\lower.5ex\hbox{$\sim$}}}}
\def\Toprel#1\over#2{\mathrel{\mathop{#2}\limits^{#1}}}

\begin{document}

\thispagestyle{empty}
\begin{flushright}
CERN-TH/2001-143\\
Parma UPRF-2001-13\\
hep-ph/0105322\\
\end{flushright}
\vspace*{15mm}

\begin{center}
{\LARGE Transverse Momentum Distributions in B-Decays \\[9pt]
}

\bigskip

$~~$\\[0pt]
{\large U. Aglietti}$^{\ast )}$ \\[0pt]
\vspace{0.1cm} $~~$\\[0pt]
Theoretical Physics Division, CERN\\[2pt]
CH - 1211 Geneva 23 \\[3pt]
$~~$ \\[0pt]
$~~$\\[0pt]
{\large R. Sghedoni$^{\ast\ast)}$} and {\large L. Trentadue$^{\ast\ast\ast)}$} \\[0pt]
\vspace{0.1cm} $~~$\\[0pt]
Dipartimento di Fisica, Universit\'a di Parma\\ and\\[2pt]
Gruppo Collegato INFN, Viale delle Scienze 7a, 43100 Parma \\[3pt]
$~~$ \\[0pt]

\bigskip {}
\end{center}
We consider transverse momentum distributions in $B$-decays. The $\cal 
O(\alpha_S)$ coefficients for soft and collinear logarithms are computed to 
next-to-leading accuracy. Resummation of
large logarithmic contributions is performed  
in impact parameter space within the general formalism for transverese 
momentum distributions. It is  shown that the shape-function 
approach as used for the threshold distribution case cannot be extended to the 
transverse momentum one. 

\vspace*{2cm}

\noindent

\bigskip

\bigskip

\bigskip

%\begin{center}
\rule[.1in]{16cm}{.002in}

%\vspace -0.2 truecm

\noindent $^{\ast )}$ On leave of absence from Dipartimento di Fisica,
Universit\'a di Roma I, Piazzale Aldo Moro 2, 00185 Roma, $~~~$ Italy.
\noindent E-mail address: {\tt ugo.aglietti@cern.ch}.\\
\noindent $^{\ast\ast)}\;\;\;$ E-mail address: {\tt roberto.sghedoni@fis.unipr.it}\\
\noindent $^{\ast\ast\ast)}\:\:\:$E-mail address: {\tt luca.trentadue@fis.unipr.it} \\

\begin{flushleft}
CERN-TH/2001-143   \\[0pt]
Parma UPRF-2001-13 \\[0pt]
May 2001
\end{flushleft}
\vfill\eject

%\clearpage\mbox{}\clearpage

%\setcounter{page}{1} \pagestyle{plain}
%\end{center}

\section{Introduction}
\pagenumbering{arabic}
In this note we discuss transverse momentum distributions in heavy flavour
decays. As may be easily recognized, the structure of the expressions 
reflects the analogous case of the energy-energy correlations or 
shape variables in $e^{+}e^{-}$ annihilation, i.e. it
involves a coefficient function, a universal function $f$ containing the
logarithmic contributions and a remainder function. This structure shows, on 
the other hand, the general form common to all the reactions such as Drell-Yan (DY), 
Deep Inelastic Scattering (DIS), Jet Shape variables, etc., once the large logarithmic 
contributions are separately resummed in the function $f$. 

The purpose of computing  the set of perturbative 
contributions for the heavy quark decay case  is introduced in this paper 
together with some preliminary results. We compute the
next-to-leading order (NLO) coefficient $B_{1}$ describing single-logarithmic effects,
which, together with the universal $A_{1}$ and $A_{2}$ coefficients,
determines the function $f$ to NLO accuracy.  

We outline here the peculiar phenomenological content of the transverse momentum 
distribution as making, for several reasons, a stand-alone case in heavy flavour physics.

The main phenomenological application of our work 
involves the spectrum of tranverse momenta of the
produced hadrons, with respect to the photon direction, in the reaction 
$b\rightarrow s\gamma .$ This process appears to be very clean because, in the rest
frame of the $B$-meson, the photon momentum unambigously fixes the reference
direction for transverse hadron momenta. 

The final goal of our study is a comparison of the complete NLO
perturbative distribution with equally accurate experimental data. 
Such a comparison should allow the
extraction of the non-perturbative component in the process i.e. the Fermi motion
effects of the $b$-quark inside the $B$-meson and the final-state 
hadronization of the strange quark. 
The information about non-perturbative dynamics
obtained with $p_{\perp }$ distributions is complementary to the one
obtained with the well-studied threshold distributions. Roughly speaking,
transverse momentum distributions are sensitive to the motion of the heavy
quark in the plane orthogonal to the decay axis, while threshold
distributions are sensitive to the motion along the decay
axis. This topic deserves some attention since, as well known, the
transverse momentum distributions of heavy flavours in hadronic processes
are presently poorly understood.

The paper is organized as follows. In sec.\thinspace 2 we present
our result on transverse momentum distributions. In sec.\thinspace
3 we summarize the main results on threshold distributions. In
sec.\thinspace 4 we discuss the results and we compare the two
different distributions. Finally, sec.\thinspace 5 contains the
conclusions together with an outlook at future developments.

\section{Transverse momentum distribution}
As anticipated in the introduction, let us consider the
distribution of relative transverse momenta between a strange
hadron $h_{s}$ and the photon in radiative $B\;$decays:
\begin{equation}
B\rightarrow h_{s}+X+\gamma ,  \label{physical}
\end{equation}
In practice, 
we expect  $h_{s}$ to be a meson such as a $K$, a $K^{\ast}$, etc.. 
Without any generality loss, we can
take the $B$-meson\ at rest and the photon\ flying along the plus direction (%
$+z$ axis); the relative tranverse momentum then coincides with the
projection of the $h_{s}$ momentum in the $x-y$ plane. In general, we may
identify three different mechanisms in transverse momentum dynamics:
\begin{enumerate}
\item  \textit{soft-gluon emission.} The elementary process in (\ref
{physical}) is: \
\begin{equation}
b\rightarrow s+\widehat{X}+\gamma .  \label{one}
\end{equation}
Non-vanishing tranverse momenta
\begin{equation}
\mathbf{p}_{p}\mathbf{\neq 0}
\end{equation}
are generated by soft-gluon emission\footnote{%
Transverse momenta are generally denoted with boldface symbols.};
\item  \textit{Fermi-motion.} The beauty quark and the light valence quark
in the $B$-meson exchange soft momenta with each other. This implies that,
before decay (\ref{one}) occurs, the $b$ quark has a non-zero transverse
momentum
\begin{equation}
\mathbf{p}_{f}\sim \Lambda
\end{equation}
where $\Lambda $ is the QCD scale.
\item  \textit{final-state hadronization.} The strange quark emitted in the
hard process fragments into the final hadron $h_{s},$%
\begin{equation}
s\rightarrow h_{s}\,+\,\left( \mathrm{anything}\right) .
\end{equation}
Since the latter is a soft process, this implies a change in the transverse
momentum of the strange system by a quantity
\begin{equation}
\mathbf{p}_{h}\sim \Lambda .
\end{equation}
\end{enumerate}
The total tranverse momentum of $h_{s}$ is then
\begin{equation}
\mathbf{p}=\mathbf{p}_{p}+\mathbf{p}_{f}+\mathbf{p}_{h}.
\end{equation}
Let us observe that Fermi motion and hadronization effects are of the same
order and cannot therefore be separated from each other. As we are going to
show later, the situation is instead different in threshold distributions.
According to the above estimates, if we take
\begin{equation}
|\mathbf{p|}\gg \Lambda ,
\end{equation}
we have that the non-perturbative effects described in 2. and 3. are
negligible, so that
\begin{equation}
\mathbf{p}\simeq \mathbf{p}_{p},
\end{equation}
and a (resummed) perturbative computation controls the distribution. The
transverse momentum distribution in impact parameter space $b$ \cite
{pp,kodtren} \ has the characteristic form \cite{cattren}--\cite{nason}:
\begin{equation}
\frac{1}{\Gamma _{B}}\frac{d\Gamma }{db}=C\left( \alpha _{S}\right)
\,f\left( b;\alpha _{S}\right) +R\,\left( b;\alpha _{S}\right) .
\label{principale}
\end{equation}
$\Gamma _{B}$ is the Born width and is given by:
\begin{equation}
\Gamma _{B}\simeq \frac{\alpha _{em}}{\pi }\frac{G_{F}\,m_{b}^{3}m_{b,%
\overline{MS}}^{2}\left( m_{b}\right) \,|V_{tb}V_{ts}^{\ast }|^{2}}{32\pi
^{3}}C_{7}^{2}\left( m_{B}\right),
\end{equation}
where we consider only the leading operator $O_{7}$ in the effective $%
b\rightarrow s\gamma $ hamiltonian \cite{misiak}.

The coefficient function $C(\alpha _{S})$ is process-dependent and short-distance
dominated and has an expansion in powers of $\alpha _{S}$ of the
form:
\begin{equation}
C\left( \alpha _{S}\right) =1+\frac{C_{F}\alpha _{S}}{\pi }c+\left( \frac{%
\alpha _{S}}{\pi }\right) ^{2}c^{\prime }+\cdots
\end{equation}
where $C_{F}=\left( N_{c}^{2}-1\right) /\left( 2N_{c}\right) =4/3,$ $\alpha
_{S}\equiv \alpha _{S}(Q^{2})$ and $Q=m_{B}$ is the hard scale. The
remainder function $R$ is process-dependent and has an expansion starting at
order $\alpha _{S}$ of the form
\begin{equation}
R\left( b;\alpha _{S}\right) =\frac{C_{F}\alpha _{S}}{\pi }r\left( b\right)
+\left( \frac{\alpha _{S}}{\pi }\right) ^{2}r^{\prime }\left( b\right)
+\cdots .
\end{equation}
$R$ vanishes in the limit of a large impact parameter:
\begin{equation}
R\left( b;\alpha _{S}\right) \rightarrow 0\quad \mathrm{for}\quad
b\rightarrow \infty ,
\end{equation}
implying that it is short-distance dominated.

The function $f\left( b;\alpha _{S}\right) $ contains the large logarithmic
contributions in a resummed form and can be written as the exponential of a
series of functions \cite{cattren2}:
\begin{equation}
f\left( b;\alpha _{S}\right) =\exp \left[ L\,g_{1}(\beta _{0}\alpha
_{S}L)+g_{2}(\beta _{0}\alpha _{S}L)+\alpha _{S}\,g_{3}(\beta _{0}\alpha
_{S}L)+\cdots \right] ,
\end{equation}
where
\begin{equation}
L\equiv \log \frac{Q^{2}b^{2}}{b_{0}^{2}}
\end{equation}
and $b_{0}\equiv 2\exp \left[ -\gamma _{E}\right] \simeq 1.12$ with $\gamma
_{E}\simeq 0.577$ the Euler constant. Our computation of the functions $g_{1}
$ and $g_{2}$ gives:
\begin{eqnarray}
g_{1}(\omega ) &=&{\frac{A_{1}}{2\beta _{0}}}\frac{1}{\omega }\left[ \log
(1-\omega )+\omega \right] , \\
g_{2}(\omega ) &=&-{\frac{A_{2}}{2\beta _{0}^{2}}}\left[ {\frac{\omega }{%
1-\omega }+}\log (1-\omega )\right] +{\frac{A_{1}\beta _{1}}{2\beta _{0}^{3}}%
}\left[ {\frac{\log (1-\omega )}{1-\omega }+\frac{\omega }{1-\omega }}+{%
\frac{1}{2}}\log ^{2}(1-\omega )\right] +{\frac{B_{1}}{\beta _{0}}}\log
(1-\omega ),\qquad
\end{eqnarray}
where
\begin{equation}
A_{1}=\frac{C_{F}}{\pi }\qquad \qquad \mathrm{and}\qquad \qquad A_{2}=\frac{%
C_{F}}{\pi ^{2}}\left[ C_{A}\left( \frac{67}{36}-\frac{\pi ^{2}}{12}\right) -%
\frac{5}{9}n_{f}T_{R}\right] .  \label{nonso}
\end{equation}
The above value of $A_{2}$ is in the $\overline{MS}$ \ scheme for the
coupling constant \cite{kodtren}.\ As usual, $C_{A}=N_{c}=3,$ $T_{R}=1/2$
and $n_{F}=3$ is the number of active quark flavours. The first two
coefficients of the $\beta $-function are:
\begin{equation}
\beta _{0}=\frac{11C_{A}-2n_{F}}{12\pi }=\frac{33-2n_{F}}{12\pi },\qquad
\qquad  \beta _{1}=\frac{17C_{A}^{2}-5C_{A}n_{F}-3C_{F}n_{F}}{24\pi
^{2}}=\frac{153-19\,n_{F}}{24\pi ^{2}}.
\end{equation}
A complete NLO analysis requires also the knowledge of the
coefficient function $c$ and of the remainder function $r\left( b\right) $
of order $\alpha _{S},$ which at present are unknown. Their computation is
in progress and will be presented in a forthcoming publication.

Denoting by $S_{1}$ the soft contribution at one loop and by $C_{1}$ the
collinear one in the notation of ref. \cite{penultimo},
\begin{equation}
S_{1}= -\frac{C_{F}}{\pi },\qquad \qquad C_{1}= -\frac{3}{4}%
\frac{C_{F}}{\pi },
\end{equation}
the next-to-leading coefficient $B_{1}$ is given by
\begin{equation}
B_{1}~=~\frac{S_{1}}{2}+C_{1}~=~ -\frac{5}{4}\frac{C_{F}}{\pi }.
\label{nonso2}
\end{equation}
In usual hard processes, such as DIS or DY, $S_{1}=0$ so that $B_{1}=C_{1}.$
In our case, single logarithmic effects are more pronounced because this
coefficient is almost a factor of 2 larger.

The expansion to order $\alpha _{S}^{2}$ of the exponent reads:
\begin{equation}
\log f\left( b;\alpha _{S}\right) =-\frac{1}{4}A_{1}\alpha
_{S}L^{2}-B_{1}\alpha _{S}L-\frac{1}{6}A_{1}\beta _{0}\alpha _{S}^{2}L^{3}-%
\frac{1}{4}A_{2}\alpha _{S}^{2}L^{2}-\frac{1}{2}B_{1}\beta _{0}\alpha
_{S}^{2}L^{2}.  \label{expandfb}
\end{equation}
Let us note that a \textit{single} constant $B_{1}$ controls the
single-logarithmic effects in any order. The physical reason is that a soft
gluon and a collinear one with the same transverse momenta are emitted with
the same effective coupling $\alpha _{S}\left( k_{\perp }^{2}\right) $ \cite
{veneziano}. This is to be contrasted with the threshold case (see next
section). The function $f(b,\alpha_S)$ is plotted in fig.~1.

\begin{center}
\psfig{bbllx=90pt, bblly=335pt, bburx=650pt, bbury=740pt,
file=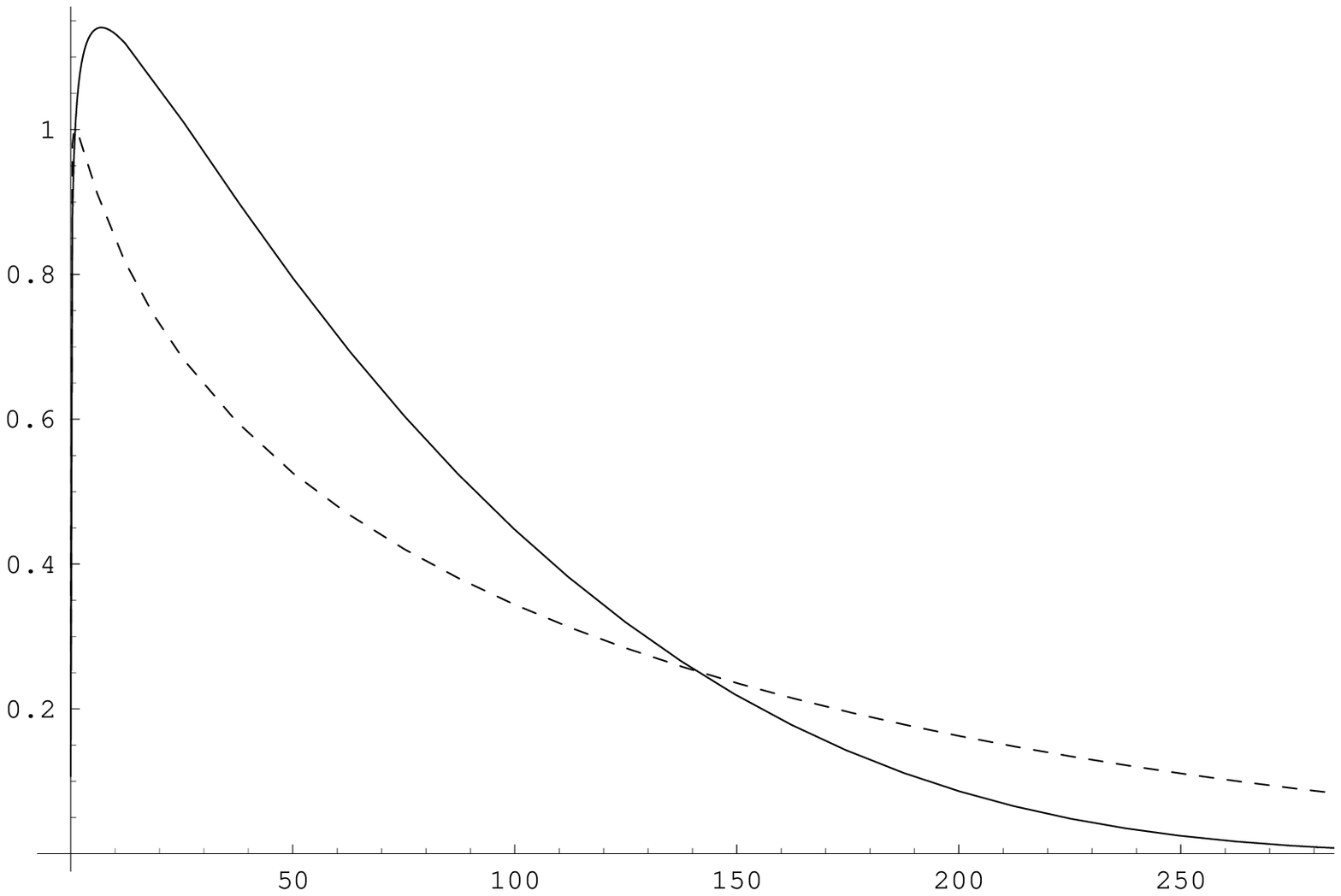, height=9cm, width=13cm} \vspace{-1.5cm}
\end{center}

\noindent \textit{Fig.~1: Plot of the function} $f(b)$ \textit{\ in the
variable }$y=Q^2 b^2 /b_0^2$ $(\alpha_S=0.22)$. \textit{\ Solid line: NLO;
dotted line: LO.}

\vspace{0.5cm}

\noindent
The resummed distribution (\ref{principale}) becomes singular when
\begin{equation}
\omega \rightarrow 1^{-}  \label{unomeno}
\end{equation}
Since
\begin{equation}
\omega =\beta _{0}\alpha _{S}L\approx \frac{\log Q^{2}b^{2}}{\log
Q^{2}/\Lambda ^{2}},
\end{equation}
the singularity occurs when the transverse strange momentum becomes as small
as the hadronic scale,
\begin{equation}
p_{\perp } \approx \frac{1}{b}\approx \Lambda .
\end{equation}
The singularity (\ref{unomeno}) is produced by the infrared pole in the
running coupling and signals an intrinsic limitation of resummed
perturbation theory, in agreement with previous qualitative analysis. Let us
note that the function $g_{2}$ has basically a pole singularity in the
limit (\ref{unomeno}), while $g_{1}$ has only a softer, logarithmic,
singularity.

\section{Threshold distribution}

In this section we recall the main results on threshold distributions. The
distribution in Mellin space --- \ the $N$-moment of the rate --- has a
similar representation to (\ref{principale}) \cite{cattren}--\cite{thrust}%
,
\begin{equation}
\frac{1}{\Gamma _{B}}\Gamma _{N}=\mathcal{C}\left( \alpha _{S}\right)
\,f_{N}\left( \alpha _{S}\right) +R_{N}\left( \alpha _{S}\right) ,
\end{equation}
where now the large logarithm contains the $N$-variable:
\begin{equation}
\qquad \qquad \qquad \qquad L\equiv \log \frac{N}{N_{0}}\qquad \qquad \left(
\mathrm{threshold\,\,case}\right) ,
\end{equation}
with $N_{0}\equiv \exp \left[ -\gamma _{E}\right] .$ The functions $g_{i}$
in the exponent are different with respect to the ones in the $p_{\perp }$
case and the leading and next-to-leading ones read \cite{americani,
penultimo2}:
\begin{eqnarray}
g_{1}\left( \lambda \right)  &=&-{\frac{A_{1}}{2\beta _{0}}}\frac{1}{\lambda
}\left[ (1-2\lambda )\log (1-2\lambda )-2(1-\lambda )\log (1-\lambda )\right]
,  \notag \\
g_{2}\left( \lambda \right)  &=&\frac{\beta _{0}A_{2}-\beta _{1}A_{1}}{%
2\beta _{0}^{3}}\left[ \log (1-2\lambda )-2\log (1-\lambda )\right] -\frac{%
\beta _{1}A_{1}}{4\beta _{0}^{3}}\left[ \log ^{2}(1-2\lambda )-2\log
^{2}(1-\lambda )\right] +  \notag \\
&&+\frac{S_{1}}{2\beta _{0}}\log (1-2\lambda )+\frac{C_{1}}{\beta _{0}}\log
(1-\lambda ).  \label{g1eg2}
\end{eqnarray}
The expansion to order $\alpha _{S}^{2}$ of the exponent reads:
\begin{equation}
\log f_{N}=-\frac{1}{2}A_{1}\alpha _{S}{L}^{2}-\alpha _{S}\left(
S_{1}+C_{1}\right) \,{L}-\frac{1}{2}A_{1}\beta _{0}\alpha _{S}^{2}L^{3}-%
\frac{1}{2}A_{2}\alpha _{S}^{2}L^{2}-\left( S_{1}+\frac{1}{2}C_{1}\right)
\beta _{0}\,\alpha _{S}^{2}\,{L}^{2}.
\end{equation}
Let us comment on the above results. The single-logarithmic effects at one
loop are controlled by the constant
\begin{equation}
S_{1}+C_{1}= -\frac{7}{4}\frac{C_{F}}{\pi },
\end{equation}
i.e. by the sum of the soft and the collinear coefficients, which is
different from the $p_{\perp }$-case (cf. eqs. (\ref{nonso2}) and (\ref
{expandfb})). At two-loop they are instead controlled by a different
constant,
\begin{equation}
S_{1}+\frac{1}{2}C_{1}=-\frac{11}{8}\frac{C_{F}}{\pi }.
\end{equation}
The soft and the collinear terms begin to differentiate at this order and
the soft one has a two times larger coefficient. Contrary to the $p_{\perp }$
case, two different constants are needed to describe the single logarithmic
effects. The dynamical difference between soft and collinear terms is that,
for a fixed jet mass, the transverse momentum of a soft gluon is
substantially smaller that of a collinear gluon \cite{veneziano,americani2,penultimo,penultimo2}.

The functions $g_{1}$ and $g_{2}$ in (\ref{g1eg2}) --- and therefore also
the resummed distribution --- have two different singularities \cite
{penultimo2,montp,lep3proc,penultimo}:

\begin{description}
\item[$i)$]  the first one occurs when
\begin{equation}
\frac{1}{2}\,=\,\lambda \,\approx \,\frac{\log Q^{2}/m^{2}}{\log
Q^{2}/\Lambda ^{2}},  \label{approx}
\end{equation}
or, equivalently, when
\begin{equation}
m^{2}\,\approx \,\Lambda \,Q,  \label{onehalf}
\end{equation}
where $m$ is the mass of the final hadronic jet $s+\widehat{X}.$ In the last
member of (\ref{approx}), we have used the approximation $N\approx
Q^{2}/m^{2}$ \cite{mangano}. The singularity (\ref{approx}) signals the
occurence of non-perturbative effects in region (\ref{onehalf}) --- to be
identified with the well-known Fermi motion \cite{nostri}; 
it is related to soft-gluon effects --- i.e. to the terms
proportional to $A_{1},\,S_{1}$ and $A_{2}$ --- and not to collinear ones
--- the term proportional to $C_{1}.$ Fermi-motion effects are therefore
controlled by soft and not by collinear dynamics. This fact allows a
factorisation of Fermi-motion effects by means of a function taking into
account soft dynamics only, the well-known shape function\footnote{
The shape function is also called structure function of the heavy flavours.} 
\cite{generale}. 
In this region, initial bound state effects become relevant while
final-state binding effects can be neglected \cite{congiulia2, penultimo,
penultimo2}.

\item[$ii)$]  the second singularity occurs at
\begin{equation}
\lambda =1,
\end{equation}
or
\begin{equation}
m^{2}\approx \Lambda \,^{2}
\end{equation}
and is related to final-state hadronization effects. Both soft and collinear
terms are singular in this region and there are non-perturbative effects
related to initial as well as final bound-state dynamics.
\end{description}

\section{Discussion and comparison of the distributions}

Let us now discuss the physics of \ $p_{\perp }$ distributions and compare
with the threshold case. In the latter case, there is a first singularity,
of soft nature, closer to the origin in $\lambda =1/2,$ which can be removed
by introducing the shape function. In the $p_{\perp }$ case, an analogous
singularity is absent, as also supported by physical intuition. In both
distributions, there is a singularity in one $(\lambda =1$ or $\omega =1),$
which cannot be removed by introducing a non-perturbative soft function.
This implis that, in the $p_{\bot }$ case, non-perturbative effects cannot
be treated with a ``shape-function'' approach. One could try a different
approach based on the effective theory introduced in \cite{UC} or the
collinear effective theory developed in \cite{separati}.

Let us note that, in general, the singularities of the functions $%
g_{i}\left( \omega \right) $ are more severe than those of the functions $%
g_{i}\left( \lambda \right) $ for $\lambda \rightarrow 1/2$ or $1.$ For
example, $g_{1}\left( \omega \right) $ has a logarithmic singularity, while $%
g_{1}\left( \lambda \right) $ has an additional prefactor $1-2\lambda $ or $%
1-\lambda $ which softens the singularity. Owing to the different
singularity structure, the $p_{\perp }$-distribution is complementary to the
threshold one and gives independent information about non-perturbative
physics.

Finally, let us make a general field-theory remark. Threshold distributions
and $p_{\perp }$-distributions have a different theoretical status. The
former are completely inclusive quantities and can be computed as the
imaginary part of a forward scattering amplitude: a non-perturbative
computation with lattice QCD is in principle feasible \cite{conguido}. This
is not the case for $p_{\perp }$-distributions, which instead are true jet
quantities: a separate computation of real and virtual diagrams is unavoidable
and a lattice QCD computation is in principle unfeasable.

\section{Conclusions}

In this note we have discussed transverse momentum distributions in
radiative $B$-decays and we have computed the next-to-leading coefficient $%
B_{1}.$ The structure of the logarithmic corrections is analogous to the one
encountered in energy-energy correlations or shape variable distributions in
$e^{+}e^{-}$ annihilations. Owing to the different singularity structure
with respect to threshold distributions, it does not seem possible to define
the analogue of a shape function. An operator definition of the
non-perturbative effects in this case should  presumably involve a different
effective theory explicitly containing the transverse degrees of freedom.

The comparison of our NLO distribution with accurate experimental data may
give new and independent information about the effective size of the
non-perturbative corrections of order $\Lambda /m_{B};$ a recurrent problem
in $B$-physics is indeed the separation of perturbative effects from
non-perturbative ones and the estimate of the latter. The computation of the
remaining next-to-leading terms --- the one loop coefficient function and
remainder function --- is in progress and will be presented elsewhere.

\begin{center}
Acknowledgements
\end{center}

One of us (U.A.) wishes to thank to S. Catani for discussions.


\begin{thebibliography}{99}
\bibitem{pp}  G. Parisi and R. Petronzio, Nucl. Phys. B 154, 427 (1979); G.
Curci, M. Greco and Y. Srivastava, Nucl. Phys. B 159, 451 (1979); Y.
Dokshitzer, D. Dyakonov and S. Troyan, Phys. Rep. 58, 269 (1980).

\bibitem{kodtren}  J. Kodaira and L. Trentadue, Phys. Lett. B 112, 66 (1982)
and preprint SLAC-PUB-2934 (1982); Phys. Lett. B 123, 335 (1983); L.
Trentadue Phys. Lett. B 151, 171 (1985); S.\thinspace Catani, E. D'Emilio
and L. Trentadue, Phys. Lett. B 211, 335 (1988).

\bibitem{cattren}  S. Catani and L. Trentadue, Nucl. Phys. B 327, 323 (1989).

\bibitem{cattren2}  S. Catani and L. Trentadue, Nucl. Phys. B 353, 183
(1991).

\bibitem{thrust}  S. Catani, L. Trentadue, G. Turnock and B. Webber, Nucl.
Phys. B 407, 3 (1993).

\bibitem{nason}  S. Frixione, P. Nason, G. Ridolfi, Nucl. Phys. B 542, 311
(1999).

\bibitem{misiak}  K. Chetyrkin, M. Misiak and M. Munz, Phys. Lett. B 400,
206 (1997); Erratum B 425, 414 (1998) and references therein.

\bibitem{penultimo} U. Aglietti, preprint CERN-TH/2001-050, hep-ph/0103002.

\bibitem{veneziano}  D. Amati, A. Bassetto, M. Ciafaloni, G. Marchesini and
G. Veneziano, Nucl. Phys. B 173, 429 (1980).

\bibitem{americani}  R. Akhouri and I.~Rothstein, Phys. Rev. D 54, 2349
(1996).

\bibitem{penultimo2}  U. Aglietti, preprint CERN-TH/2001-035, hep-ph/0102138.
  
\bibitem{americani2}  G.~Korchemsky and G. Sterman, Phys.\thinspace
Lett.\thinspace B 340, 96 (1994).

\bibitem{UC}  U. Aglietti and G. Corbo, Phys. Lett. B 431, 166 (1998) and
Int. J. Mod. Phys. A 15, 363 (2000).

\bibitem{montp}  U. Aglietti, Nucl. Phys. B Proc. Suppl. 96, 453 (2001).

\bibitem{lep3proc}  U. Aglietti, talk given at the LEP3 conference, Rome
18-20 April 2001, to be published in the proceedings, hep-ph/0105168.

\bibitem{mangano}  S. Catani, M. Mangano, L. Trentadue and P. Nason, Nucl.
Phys. B 478, 273 (1996).

\bibitem{nostri}  G. Altarelli, N. Cabibbo, G. Corb\'{o}, L. Maiani and G.
Martinelli, Nucl. Phys. B 208, 365 (1982).

\bibitem{generale}  I. Bigi, M. Shifman, N. Uraltsev and A. Vainshtein,
Phys. Rev. Lett. 71, 496 (1993); Int. J. Mod. Phys. A 9, 2467 (1994); A.
Manohar and M. Wise, Phys. Rev. D 49, 1310 (1994); M. Neubert, Phys. Rev. D
49, 3392 and 4623 (1994); T. Mannel and M. Neubert, Phys. Rev. D 50, 2037
(1994).

\bibitem{congiulia2}  U. Aglietti and G. Ricciardi, Nucl. Phys. B 587, 363
(2000).

\bibitem{separati}  C. Bauer, S. Fleming and M. Luke, Phys. Rev. D 63,
014006 (2001); C. Bauer, S. Fleming, D. Pirjol and W. Stewart, Phys. Rev. D
63, 114020 (2001).

\bibitem{conguido}  U. Aglietti, M. Ciuchini, G. Corb\'{o}, E. Franco, G.
Martinelli and L.~Silvestrini, Phys. Lett. B 432, 411 (1998).
\end{thebibliography}
\end{document}